# Direct Determination of Band Gap Renormalization in Photo-Excited Monolayer MoS$_2$

Fang Liu, Mark Ziffer, Kameron R. Hansen, Jue Wang, Xiaoyang Zhu[*]

Department of Chemistry, Columbia University, New York, NY 10027, USA

**A key feature of monolayer semiconductors, such as transition-metal dichalcogenides, is the poorly screened Coulomb potential, which leads to large exciton binding energy ($E_b$) and strong renormalization of the quasiparticle bandgap ($E_g$) by carriers. The latter has been difficult to determine due to cancellation in changes of $E_b$ and $E_g$, resulting in little change in optical transition energy at different carrier densities. Here we quantify bandgap renormalization in macroscopic single crystal MoS$_2$ monolayers on SiO$_2$ using time and angle resolved photoemission spectroscopy (TR-ARPES). At excitation density above the Mott threshold, $E_g$ decreases by as much as 360 meV. We compare the carrier density dependent $E_g$ with previous theoretical calculations and show the necessity of knowing both doping and excitation densities in quantifying the bandgap.**

Atomically thin transition-metal dichalcogenide (TMDC) monolayers and heterojunctions are being broadly explored as model systems for a wide range of electronic, optoelectronic, and quantum processes. The commonly studied TMDC monolayers possess direct bandgaps in the visible to near-IR region [1–3]. Because of the strong many-body Coulomb interactions in monolayer TMDCs, both exciton binding energy ($E_b$) and bandgap renormalization energy are large [3]. The former lowers the optical transition energy by hundreds meV from $E_g$, while the latter decreases $E_g$ by similar amounts in the presence of charge carriers or excitons. The bandgap renormalization energy ($\Delta E_g$) and decrease in exciton binding energy ($\Delta E_b$) tend to be of similar magnitudes but counteract each other, leading to comparatively modest changes in optical transition energies [4,5]. Since the quasiparticle bandgap $E_g$ is the most fundamental quantity and is predicted to be exceptionally sensitive to carrier or exciton densities [4,6,7], there is clearly a need to determine bandgap renormalization and its dependence on carrier/exciton densities.

---

[*] To whom correspondence should be addressed. E-mail: xyzhu@columbia.edu

Past attempts at measuring $\Delta E_g$ required analysis of subtle or small features in optical spectra [8–11]. Examples include estimating the gain threshold in transient reflectance spectra from photo-excited TMDC monolayer and bilayer above the Mott density [8], extrapolating $E_g$ from the experimental Rydberg exciton series in conjunction with theoretical models [9,10], and identifying features attributed to bandgap transition on the broad fluorescence excitation spectra of gate-doped monolayer $MoS_2$ [11]. The ideal technique to determine quasiparticle energies is angle resolved photoelectron spectroscopy (ARPES), which directly maps band energies with momentum resolution. ARPES typically probes the valence bands and populating the conduction band would require either a) heavy chemical doping via K or H atom deposition [12–15] or b) photo-doping via above-gap optical excitation in time-resolved (TR) ARPES [16–18]. The chemical doping approach may lead to undesirable changes to the dielectric environment and lattice structure of TMDC monolayers [12–15,19]. TR-ARPES of transiently excited TMDCs can in principle probe both the quasiparticle bandgap and the dynamics of bandgap renormalization. However, past attempts of TR-ARPES on TMDC monolayers have used CVD grown polycrystalline monolayers on metal or semimetal substrates. These conductive substrates drastically modify both the energetics and dynamics of excited states in TMDC monolayers [16–18]. Others TR-ARPES studies have used bulk TMDC crystals, instead of monolayers [20–22]. To overcome these limitations, here we prepare single crystal $MoS_2$ monolayers with macroscopic sizes (mm-cm) on dielectric substrates (285 nm thick $SiO_2$ on n-doped Si). We use TR-ARPES to monitor the time evolution of the valence band maximum (VBM) and conduction band minimum (CBM) following above-gap optical excitation. We directly quantify bandgap energies with excitation density across the Mott threshold and compare experimental results with recent theoretical calculations.

In our femtosecond TR-ARPES experiment (Fig. 1a and Fig. S3), the visible excitation pulse ($h\nu_1$ = 2.2 eV, 40 fs pulse width, s-polarized) is obtained from a home-built non-colinear optical parametric amplifier (NOPA), pumped by a Ti:Sapphire laser (Coherent Legend, 10 W, 10 KHz, 800 nm, 35 fs). Part of the Ti:Sapphire laser output is frequency-doubled for high harmonic generation (HHG) in Kr gas (KM Labs, XUUS) to produce EUV probe pulses ($h\nu_2$ = 22 eV, pulse duration <100 fs, p-polarized) [23]. The EUV pulse ionizes electrons from both valence and conduction bands for detection by a hemispherical analyzer with angular resolution. Note that the use of EUV, instead UV probe, is necessary to access the high momentum K point at the Brillouin zone corner. The plane of light incidence and analyzer slit is parallel to the *Γ-K* direction, with the sample azimuthal geometry fixed for the collection of photoemission from the K valley.

Fig. 1b shows optical image of a single crystal $MoS_2$ monolayer (blue color) on the $SiO_2$/Si substrate. Our improvement to the gold exfoliation technique [24] has yielded macroscopic single crystal samples (lateral dimension in the mm-cm range). See Fig. S1 and S2 in Supporting Materials for characterization by atomic force microscope (AFM) for sample cleaness and second harmonic generation (SHG) for alignment. The complex dielectric function (Fig. 1c) obtained from white light reflection shows the characteristic A and B excitons and photoluminescence spectrum (Fig. 1d) shows emission from the A exciton ($E_A$ = 1.865±0.05 eV). These optical spectra are consistent with those of previous reports [1,25].

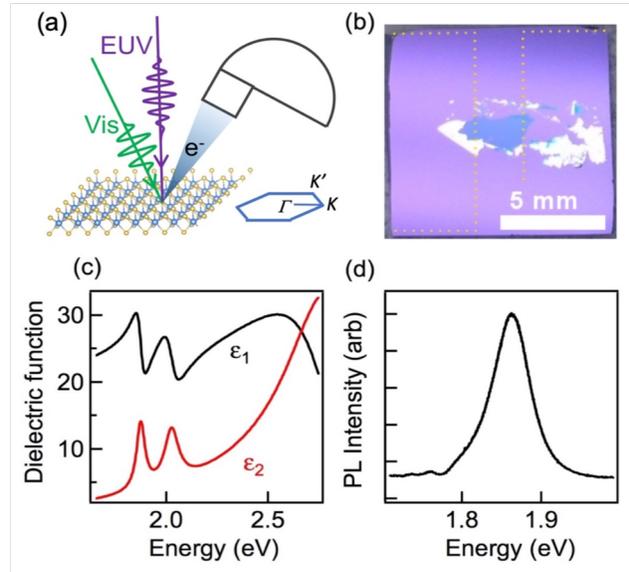

**Figure 1**. **The macroscopic single crystal MoS2 monolayer sample and characterization.** (a) Schematics of TR-ARPES experiment, combing the femtosecond visible pump and EUV probe. The photoelectrons are collected by the hemispherical analyzer at a specific angle θ from the surface normal, corresponding to emission from a K valley. (b) Image of the single crystal $MoS_2$ monolayer. We deposit Au films in the dashed areas for electrical contact and grounding. (c) complex dielectric function ($\varepsilon=\varepsilon_1+i\varepsilon_2$) of the monolayer $MoS_2$ determined from reflection spectroscopy and (d) Photoluminescence of the $MoS_2$ monolayer at room temperature.

In a TR-ARPES experiment, the visible pump pulse induces a direct transition in the K and K' valleys. Following a controlled time delay (Δt), the EUV probe pulse ionizes the electrons in the valence and conduction bands. Fig. 2 shows momentum-resolved ARPES from monolayer $MoS_2$ around the K valley without (a) and with (b) visible pump (Δt = 0). The two spectra are integrated over the 1.1-1.4 Å$^{-1}$ parallel momentum window to yield the corresponding energy distribution curves (EDCs), shown in Fig.2(c). The MoS$_2$ monolayer sample is n-doped, with the Fermi energy close to the CBM. As a result, weak photoelectron signal from intrinsic population in the conduction band near CBM is observed in Figure 2a. This signal is used to determine a doping density of $n_0 = (4.9\pm1.0) \times 10^{12}$ cm$^{-2}$ (see Supporting Information, Figs. S5 and S6). Mechanically exfoliated MoS$_2$ monolayers are commonly known to be of n-type at similar doping levels [26–28].

With the addition of the pump pulse, photo-excitation across the bandgap creates exciton and/or electron/hole carrier density, $n_{e/h}$, on top of $n_0$. In the experiment, we vary the excitation densities $n_{e/h}$ in the range of $3.8\times10^{12}$ to $2.3\times10^{13}$ cm$^{-2}$. For reference, the Mott density for the transition from exciton gas to electron-hole plasma is $n_{Mott} \sim 4.3\times10^{12}$ cm$^{-2}$, estimated from the 2D scaling relationship [7] of $a_0 n_{Mott}^{1/2} \approx 0.25$ with $a_0$ (exciton Bohr radius) = 1.2 nm for undoped monolayer MoS$_2$ on SiO$_2$ [29]. The excitation density probed here is mostly in the e-h plasma region. Fig. 2b shows ARPES spectra at excitation density of $n_{e/h} = (2.3\pm0.5) \times 10^{13}$ cm$^{-2}$ at $\Delta t = 0$. Compared to Fig. 2a, we observe three major changes: 1) an increase in conduction band electron intensity; 2) an up-shift in the VBM; and 3) a broadening in the valence band EDC. Photoelectron signal from the conduction band at $\Delta t = 0$ probes $n_0+ n_{e/h}$; therefore the prompt increase in CB photoemission signal, observation 1), is proportional to the excitation density $n_{e/h}$. To understand observations 2) and 3), we point out that the depletion in the valence band by optical excitation is ~1% of the total electron density in the band and not detectable in our experiment. Thus, the up-shift in VBM and broadening of the valence band results from manybody effects resulting from the excitation. The former measures the band renormalization [4,6,7] and the latter attributed to dephasing from hole-hole scattering [30].

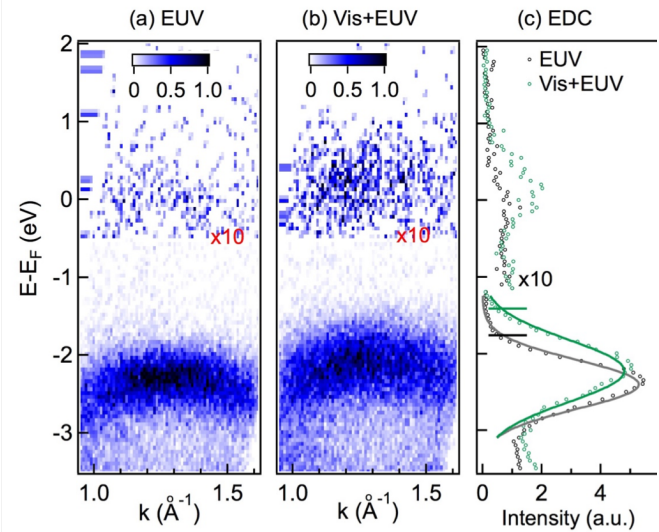

**Figure 2**. **TR-ARPES from monolayer MoS$_2$.** (a) (b) EUV ARPES of single crystal MoS$_2$ monolayer without and with the visible pump excitation ($\tau=0$). The APRES spectra is collected at K valley along Γ to K direction. The visible pump is at a photon energy of 2.2 eV. (c) Corresponding electron energy distribution curves (EDC). The solid lines are Gaussian functional fits. The horizontal marks represent the edges of the EDCs, corresponding to $E_0+2\sigma$. The spin orbit splitting at K valley is not resolved under the current energy resolution. The excitation density from the pump pulse is $n_{e/h} = (2.3\pm0.5) \times 10^{13}$ cm$^{-2}$. The conduction band signal is magnified by 10x for clarity.

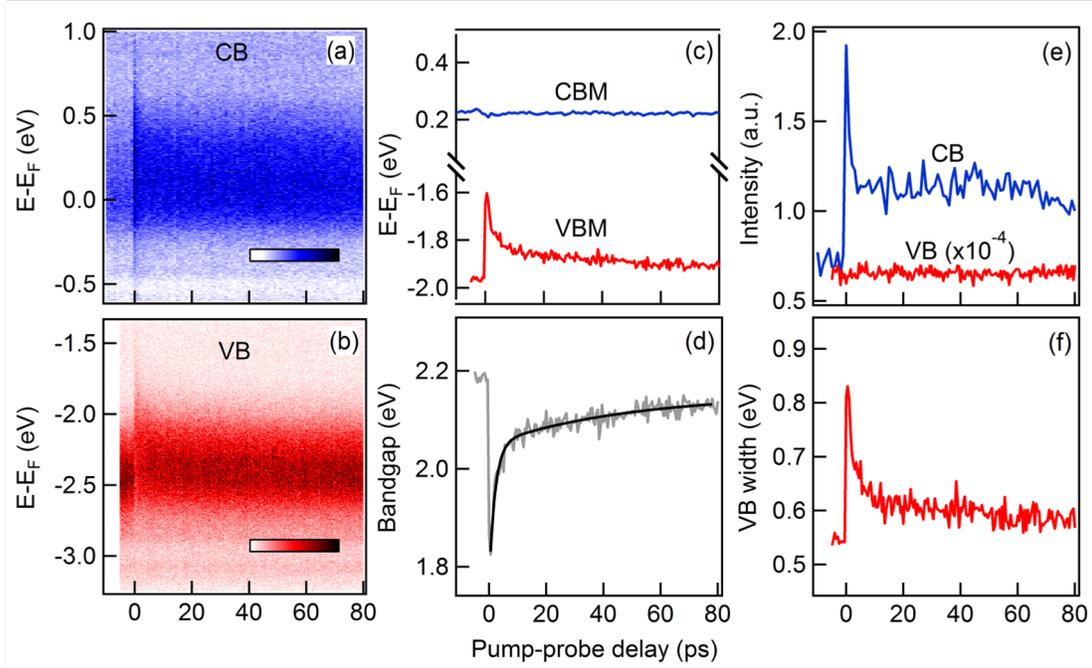

**Figure 3. Dynamics of band renormalization.** All panels are shown as a function of pump-probe delay: 2D pseudo-color (intensity) plots of EDC spectra of conduction band (a) and valence band (b); (c) CBM and VBM positions, (d) band gap $E_g$ (grey) with bi-exponential fit (black solid line); (e) conduction (blue) and valence (red) band photoelectron intensities; and (f) full width at half maximum of valence band. To obtain the EDCs in (a) and (b), the photoelectron signal is integrated from 1.1 Å$^{-1}$ to 1.4 Å$^{-1}$. The CBM is fixed at the time averaged value of 0.223 eV in the calculation of the bandgap in (d). The initial excitation density is $n_{e/h}$ = 1.3 x 10$^{13}$ cm$^{-2}$ and sample is at 295 K. The color scales in (a) and (b) are normalized (0-1).

We now turn to the dynamics of the manybody effects following optical excitation. Figs. 3a and 3b are 2D pseudo-color EDC plots showing the conduction band and valence band photoemission signal, respectively, as a function of pump probe delay Δt. Because of the low electron population in the conduction band, we assume these electrons reside close to the CBM and take the intensity-weighted average of the CB photoelectron energies as the CBM position. For the VBM, a common practice in photoemission studies is to use linear extrapolation near the threshold, which may introduce large uncertainty. Instead, each valence band EDC from the K valley is well described by a Gaussian function, and therefore we use the high energy cutoff at $E_a+2\sigma$ ($E_a$ is the intensity-weighted average of the valence band energy and $\sigma$ is variance of the Gaussian fit) to represent the VBM. These two approaches yield similar VBM values, as shown in Supporting Information (Fig. S5).

Fig. 3c show the VBM/CBM positions as a function of $\Delta t$. The difference between CBM and VBM gives the time-dependent $E_g$ shown in Fig. 3d. $E_g$ is measured to be 2.19 ± 0.10 eV in the absence of optical excitation ($\Delta t < 0$), which is ~0.4 eV lower than $E_g$ = 2.6±0.2 eV in undoped monolayer $MoS_2$ [6,11]. This difference reflects band renormalization from the intrinsic n-type doping of $n_0$ = 4.9±1.0 x $10^{12}$ $cm^{-2}$ [6,11]. At $\Delta t$ = 0, photoexcitation across the bandgap further lower $E_g$ by as much as $\Delta E_g$ = -0.36 ± 0.04 eV. Interestingly, the photo-excitation induced bandgap renormalization is reflected exclusively in the up-shift in the VBM while the CBM remains constant, suggesting that the CBM is pinned to the Fermi-level of the metal contact in our n-doped sample, in agreement with Bampoulis et al. [27].

The large photo-induced bandgap renormalization results from the poorly screened Coulomb potential and strong many body interactions in the TMDC monolayer. The renormalized bandgap, initially by $\Delta E_g$ = -0.36 ± 0.04 eV at $\Delta t$ = 0, recovers with increasing $\Delta t$ due to carrier recombination. This recovery can be described by a bi-exponential fit (solid curve in Fig. 3d), with time constants of $\tau_d$ = 2 ps and 80 ps, respectively. The time-dependence in $E_g$ is consistent with the population decay of conduction band photoelectron intensity (blue curve in Fig. 3e), as well as in the recovery of valence band width (Fig. 3f). For comparison, the photoelectron intensity from the valence band (red curve in Fig. 3e) remains constant, as expected from the small depletion of the valence band (~1%) due to photo-excitation. At such a high excitation density, the fast decay ($\tau_d$ = 2 ps) likely results from Auger recombination [32], while the slow-decay can be attributed to intrinsic radiative/nonradiative decays in the $MoS_2$ monolayer [33]. We point out that electronic interaction with or screening by the $SiO_2$ dielectric substrate is minimal for our single crystal $MoS_2$ monolayer and photo-excited carrier populations survive for over 400 ps at room temperature (see Fig. S9). For comparison, previous experiments of polycrystalline TMDC monolayers on metal or graphene substrates show lifetimes up to four-orders of magnitude shorter [17,18]. Thus, the band renormalization determined here reflects close-to intrinsic manybody interactions in the $MoS_2$ monolayer [30]. The time dependent band renormalization quantified in our TR-ARPES measurement is also in qualitative agreement with previous optical measurements on monolayer TMDCs [7,8,11,35–37].

Our ability to determine bandgap renormalization in macroscopic MoS$_2$ monolayers on an dielectric substrate allows us to carry out quantitative comparisons to theoretical predictions [4,6,7,38]. The solid circles in Fig. 4 are bandgap values determined in our TR-ARPES measurements at different excitation densities. The photoinduced electron/hole density $n_{e/h}$ coexists with the intrinsic electron density of $n_0 = 4.9\pm1.0 \times 10^{12}$ cm$^{-2}$ from n-type doping. Both e/h pairs from optical excitation and excess electrons from n-type doing screen the Coulomb interactions, leading to renormalization of the bandgap. We use the total carrier density, i.e., $n_0+2n_{e-h}$, in comparison to previous calculations for either electron doping or e/h pair excitation. Liang et al. calculated band renormalization (red curve) from the quasiparticle self-energies of valence and conduction bands in monolayer MoS$_2$ using a new plasmon-pole model that takes into account carrier occupation and carrier screening at high electron doping levels [6]. The result is close to that of a more recent GW calculation (blue curve) by the same group [38]. Mechbach et al. incorporated plasma dielectric screening into a four band Hamiltonian and solved the Dirac-Bloch equation to obtain renormalized band gaps at different excitation densities ($n_{e/h}$) [7]. The resulting bandgap values (green curve) is below those predicted for only electron doping (red and blue curves). While the first data point for $n_0 = 4.9\pm1.0 \times 10^{12}$ cm$^{-2}$ in our measurement (without photo-excitation) is very closer to the theoretical results for the same electron doping density [6,38], the experiment data points move closer to the results of Mechbach et al. for photo-doping [7]. This comparison reveals the critical importance of knowing both intrinsic doping levels and additional photo-excitation densities in quantifying the bandgap in 2D TMDCs.

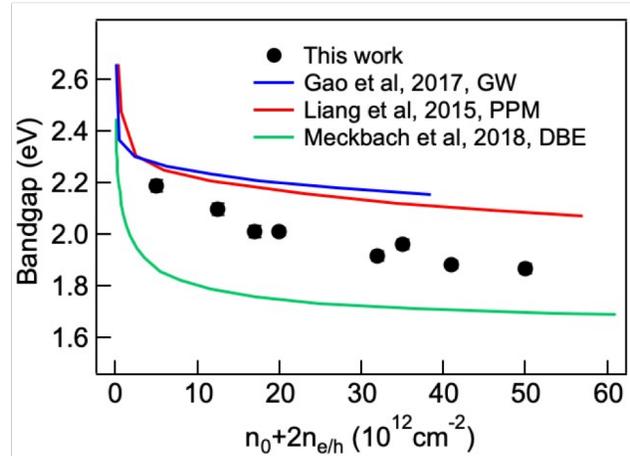

**Figure 4. Comparison of experimental bandgaps with theoretical calculations.** Solids circles are experimental bandgaps for monolayer MoS$_2$ determined by TR-ARPES as a function of conduction band electron density ($n_0+n_{e/h}$). The solid curves are theoretical results for electron doping (red [6] and blue [38]) or e/h pair generation from optical excitation (green) [7].

In summary, we carry out direct and quantitative measurement of bandgap renormalization in photo-excited $MoS_2$ monolayers using TR-ARPES. The use of macroscopic and single crystal $MoS_2$ samples on a dielectric ($SiO_2$) surface allows us to access the close-to-intrinsic bandgap renormalization and carrier decay dynamics in the 2D semiconductor. We show reduction in the bandgap by as much -0.36 eV for photo-excitation above the Mott density in an n-type $MoS_2$ monolayer. The measured density-dependent bandgap provides a benchmark for the validation of theoretical models and for the understanding of strong manybody interactions in TMDC monolayers.

**Supporting Information**: Experimental methods, including 1) the preparation of macroscopic $MoS_2$ single crystal monolayer; 2) characterization by reflectance spectroscopy and photoluminescence spectroscopy; and 3) time-resolved ARPES measurements. Additional data and analysis.

**Acknowledgement**

XYZ acknowledges the National Science Foundation (NSF) grant DMR-1608437 for supporting the TR-ARPES measurements and NSF grant DMR-1809680 for supporting the sample preparation. XYZ acknowledges the Center for Precision Assembly of Superstratic and Superatomic Solids, a Materials Science and Engineering Research Center (MRSEC) through NSF grant DMR-142063 for supporting the purchase and development of the extreme-UV laser source and for the optical characterization of monolayer $MoS_2$ samples. Partial supports by the Columbia Nano Initiative and the Vannevar Bush Faculty Fellowship through Office of Naval Research Grant # N00014-18-1-2080 for the purchase of the laser equipment are also acknowledged. FL acknowledges support by the Department of Energy (DOE) Office of Energy Efficiency and Renewable Energy (EERE) Postdoctoral Research Award under the EERE Solar Energy Technologies Office administered by the Oak Ridge Institute for Science and Education (ORISE). ORISE is managed by Oak Ridge Associated Universities (ORAU) under DOE contract number

DE-SC00014664. All opinions expressed in this paper are the author's and do not necessarily reflect the policies and views of DOE, ORAU, or ORISE. We acknowledge Yusong Bai for help in SHG and PL experiments.

SUPPLEMENTARY MATERIAL

**Direct Determination of Band Gap Renormalization in Photo-Excited Monolayer MoS$_2$**

Fang Liu, Mark Ziffer, Kameron R. Hansen, Jue Wang, Xiaoyang Zhu[*]

Department of Chemistry, Columbia University, New York, NY 10027, USA

## 1. Experimental

### Sample preparation

We prepare single crystal monolayer MoS$_2$ samples with mm-cm lateral sizes using a gold mediated exfoliation technique reported by Desai et al. [1]. Briefly, a 150 nm thick gold thin film is thermally evaporated onto a bulk MoS$_2$ single crystal (SPI Supplies). The topmost MoS$_2$ layer that strongly bond with gold is peeled off from the bulk crystal, together with the gold layer, by a thermal release tape (Semiconductor Corp.) and is subsequently transferred onto a silicon substrate with a 285 nm thick SiO$_2$ layer. The thermal release tape is removed at 120 °C, leaving behind the gold covered MoS$_2$ monolayer on the silicon substrate. The sample is cleaned with O$_2$ plasma and acetone to remove tape residues. The gold film is etched away with a concentrated aqueous solution of KI and I$_2$ (≥99.99%, Alfa Aesar), leaving behind a single crystal MoS$_2$ monolayer on the silicon dioxide. Atomic force microscope (AFM) imaging confirms an atomically flat and clean surface (Fig. S1). For grounding in photoemission experiments, a 150 nm thick gold film is evaporated onto two sides of the macroscopy MoS$_2$ monolayer through a shadow mask (see Fig. 1b in the main text). The MoS$_2$ monolayer with lateral dimension of ~2 mm between the two gold contacts is used in photoemission experiment. We determine the orientation of the MoS$_2$ monolayer using second harmonic generation (SHG) imaging, as shown in Fig. S2. The orientation of the sample is set so that the plane of light incidence and photoelectron detection is in the Γ-K direction of the MoS$_2$ monolayer.

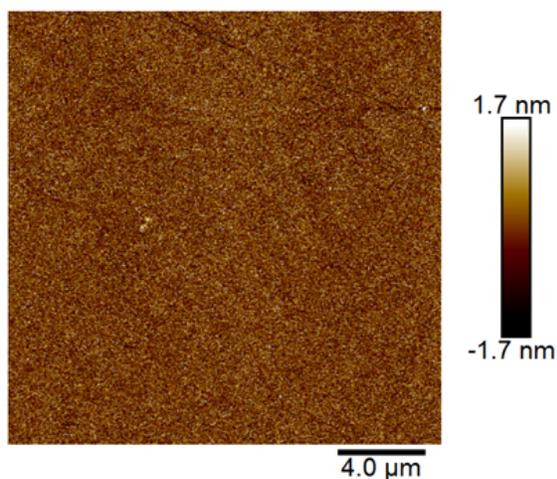

Figure S1. Atomic Force Microscope (AFM) iamge of the monolayer MoS$_2$ supported on SiO$_2$/Si. Measurement carried out on a Bruker Dimension FastScan AFM using Peakforce tapping mode.

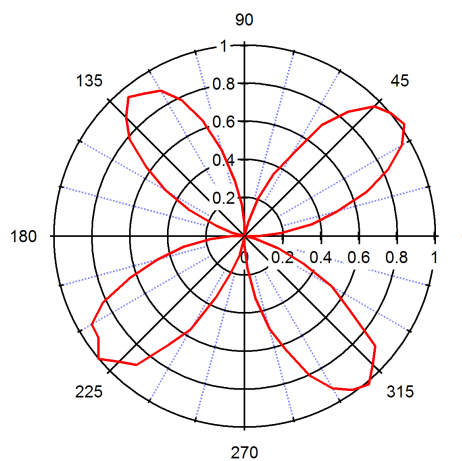

Figure S2. Determination of MoS$_2$ monolayer Γ-K orientation via polarization-resolved SHG, via rotating input laser polarization. The polarization of laser is rotated while the sample and SHG detection polarization are fixed. The SHG exhibit a four-fold symmetry when plotted against the rotation angle of the polarizer. The correlation between the phase and Γ-K (zigzag) direction is calibrated with fit from SHG data obtained from CVD grown WS$_2$ monolayers. For 2PPE measurements, the sample is positioned with Γ-K direction parallel to the detection slit.

[*] To whom correspondence should be addressed. e-mail: xyzhu@columbia.edu

Time-resolved and angle-resolved photoemission spectroscopy (TR-ARPES)

The MoS$_2$ monolayer sample supported on the SiO$_2$ substrate is transferred to an ultrahigh vacuum chamber (~10$^{-10}$ Torr) and annealed at 200 °C for 4 h via direct current heating to remove residual contaminants. The TR-ARPES measurements are carried out in the ultrahigh vacuum at room temperature. The kinetic energy of the photoemitted electrons are measured on a hemispherical electron energy analyzer equipped with a 2D delay line detector (SPECS Phoibos-100).

For TR-ARPES, a home-built noncolinear optical parametric amplifier (NOPA) output in the visible region and the EUV output from high harmonic generation are used as pump and probe, respectively, Fig. S3. In particular, the output of a regenerative Ti:Sapphire amplifier (Coherent Legend Elite Duo HE+, 10 kHz, 800 nm, 10 W, 35 fs) is split into two branches with a 80/20 beam-spliter. 20% of the output is frequency converted by NOPA, which generates output tunable between 500-650 nm, with typical pulse duration of 35 fs and average power of 10-20 mW. The remaining 80% of the output from the Ti:Sapphire amplifier is sent to a BBO crystal (Laserton) and is frequency doubled to 400 nm at an average power of 3 W. The residual fundamental 800 nm is separated from the 400 nm with dichroic mirrors. The 400 nm light is focused into a Kr gas filled hollow fiber in an ultrafast extreme ultraviolet (EUV) setup (KMLabs, XUUS4, modified for 400 nm pump). The high harmonic generation in Kr results in EUV in odd harmonics of the 400 nm pump, peaking at 22 eV (7$^{th}$ harmonic). The output EUV beam is separated from the 400 nm pump by a Si mirror and refocused with a gold coated toroidal mirror downstream. The lower harmonics and the remaining 400 nm are blocked by a 100 nm Al filter before it reaches onto the sample in the vacuum chamber. Photoemission replicas from harmonics other than the 7$^{th}$ are not observed in experimental photoemission spectra due to their low intensities. The visible pump beam is s-polarized while the probe EUV is p-polarized. The energy resolution is on the order of 300 meV and the time resolution is on the order of 50 fs. The sample is kept at room temperature in TR-ARPES measurements.

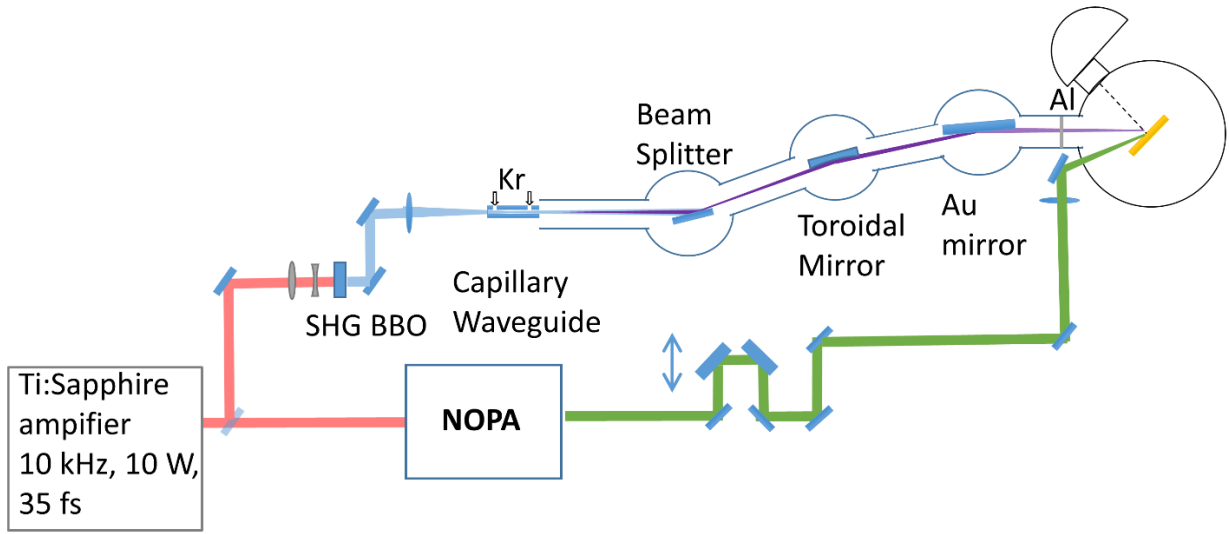

Figure S3. EUV generation and the time-resolved ARPES setup. The capillary waveguide, beam-splitter, toroidal mirror and gold mirror are all under vacuum.

Photoluminescence and reflectance spectra

All optical measurements of the sample were carried out in a $N_2$ gas filled cell at room temperature. The steady photoluminescence (PL) spectrum was collected on an InVia Raman Microscope (Renishaw), with 532 nm excitation. The reflectance spectrum of the $MoS_2$/ $SiO_2$(285nm)/Si sample was measured on a custom set-up built around an inverted optical microscope (Nikon TE300). Unpolarized white light from a tungsten-halogen source (Fostec 8375 with Thorlabs OSL2BIR lamp) was focused onto a pinhole and collimated by a 0.7 NA lens (f =

16 mm). The collimated white light was reflected off a 50/50 beam splitter (Thorlabs BSW29) and brought into focus on the sample plane by a 40X objective (Nikon S Plan Fluor ELWD). The focused spot size was roughly 3 μm. The light reflected off the sample was collected via transmission through the 50/50 beam splitter and imaged onto the entrance slit of a spectrograph (Princeton Instruments SpectroPro HRS-300) coupled to a Si EMCCD camera (Princeton Instruments Photon Max 512) to record the reflected intensity ($I_{reflected}$) spectrum. Dark spectra were measured with the white light blocked in order to account for detector dark counts. The incident intensity was determined by measuring the reflected intensity from a bare $Si/SiO_2$(285nm) substrate, which was then spectrally corrected based on the known reflectance spectrum for the substrate to give a corrected incident intensity spectrum ($I_{incident}$). Reflectance of the full $Si/SiO2$(285nm)/$MoS2$ stack was then calculated according to $R=I_{reflected}/I_{incident}$.

Transfer matrix modelling was used to determine the complex dielectric function ($\epsilon = \epsilon_1 + i\epsilon_2$) of $MoS_2$ by fitting a simulated reflectance spectrum to the experimental reflectance spectrum of the $Si/SiO_2$(285nm)/$MoS_2$ stack. The thickness of the $MoS_2$ layer for transfer matrix modelling was fixed at 0.615 nm according to literature value [2]. As an initial guess for the dielectric function, an oscillator model using a sum of five Lorentzian oscillators was first parameterized to fit the $\epsilon_1$ and $\epsilon_2$ spectra of $MoS_2$ reported by Li *et al.* [2]. The corresponding parameters of the oscillator model were then used as an initial guess to iteratively fit the simulated reflectance spectrum of the stack to the experimental spectrum using the transfer matrix method. The $\epsilon_1$ and $\epsilon_2$ spectra according to the best fit oscillator model are shown in Fig 1c. of the main text, and were also used to simulate the fraction of light absorbed by the $MoS_2$ monolayer in the $Si/SiO_2/MoS_2$ stack (Fig S4) using the transfer matrix method.

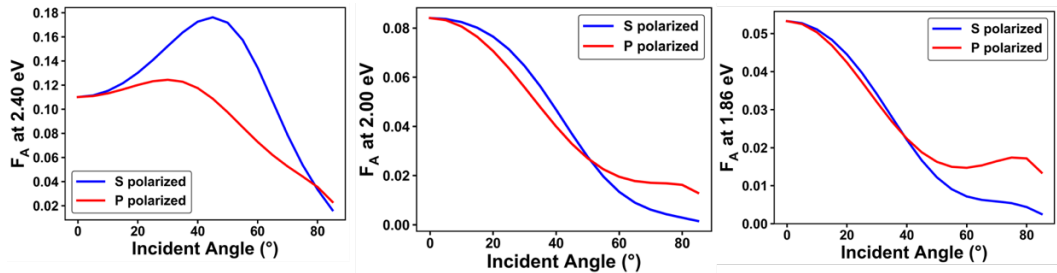

Figure S4. Simulated fraction of light absorbed ($F_A$) by the $MoS_2$ monolayer in the $MoS_2/SiO_2/Si$ stack as a function of incident angle and linear polarization (S vs. P) for 3 different absorption energies (2.4 eV, 2.0 eV, 1.86 eV) using the dielectric function (Fig. 1c) and transfer matrix modelling.

## 2. Additional data and analysis

Determination of valence band maximum

We determine the valence band maximum (VBM) based on the energy distribution curve (EDC) from the K valley. The EDC is well described by a Gaussian function; we use the high energy cutoff at $E_a+2\sigma$ ($E_a$ is the intensity-weighted average of the valence band energy and $\sigma$ is variance of the Gaussian fit) to represent the VBM. As shown in Fig. S5, the VBM position determined with this method (vertical line) is in excellent agreement with that from the common practice of linear extrapolation (dashed lines) near the threshold. The VBM determined with the $E_a+2\sigma$ is more reproducible than that from the linear extrapolation.

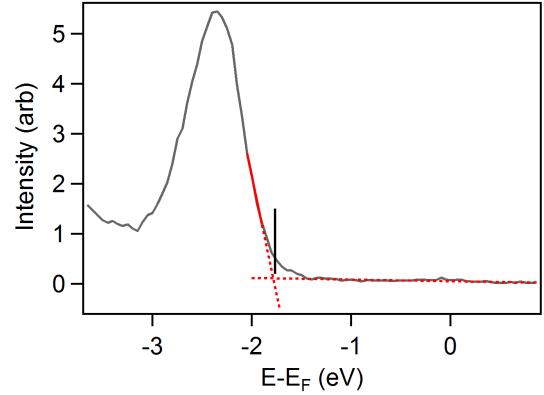

Figure S5. EDC of the EUV photoemission of monolayer $MoS_2$ at K point (grey curve). The vertical bar represents $E_0+2\sigma$ in the Gaussian fit shown in Figure 2. The red dashed lines are linear extrapolation of the baseline and the fall off edge. The crossing of the two lines are conventionally used to obtain band edge position.

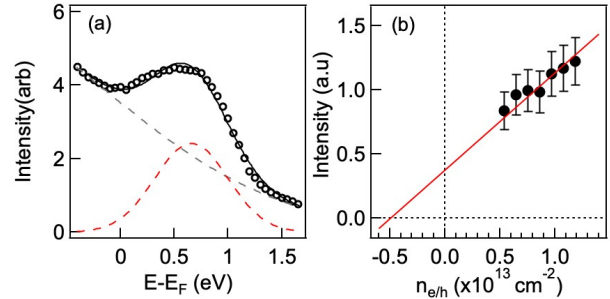

Determination of excitation density $n_{e/h}$

The injected carrier density $n_{e/h}$ per laser pulse (cm$^{-2}$) is determined by

$$n_{e/h} = \frac{P \cdot A}{S \cdot f_{\text{rep}} \cdot h\nu}$$

where $P$ and $S$ are the power (W) and cross-sectional (cm$^2$) area of the incident laser, $f_{\text{rep}}$ (= 10 kHz) is the pulse repetition rate, $h\nu$ is the photon energy (J), $A$ is the absorbance of the sample determined from the dielectric function (See Fig. S4). While the absorption cross section may deviate from the linear relationship at sufficiently high excitation densities, we find this deviation is negligible in the density range used here (see Fig. S6b).

Figure S6. Determination of intrinsic doping density. (left) Laser excitation density and corresponding measured CB electron density increase dn divided by the CB density $n_{t<0}$. The fitting slope is 1.30 x 10$^{13}$ cm$^{-2}$. This slope corresponds to added intensities of CB electrons + baseline. (Right) Extraction of the CB intensity from the baseline. The EDC is fitted with a Gaussian peak (red dashed line) on top of a Gaussian tail baseline (grey dashed line). The ratio of the CB electron density / baseline is therefore determined. The intrinsic doping density is obtained to be (4.92±0.16) x10$^{12}$ cm$^{-2}$.

Determination of intrinsic doping density

The EDC (circles in Fig. S6a) of conduction without pump excitation can be fitted with a Gaussian peak (red dashed curve in Fig. S6a) on top of a Gaussian tail baseline; the latter is background signal. We use the integrated Gaussian peak as a representation of the conduction band electron density, solid circles in Fig. S6b plotted against the photo excitation density ($n_{e/h}$). A linear fit (red line) of photoelectron signal versus gives an x-offset at $-4.9 \pm 1.0 \times 10^{12}$ cm$^{-2}$; note that the error bar mainly results from the uncertainty in the determination of excitation density. Thus, the intrinsic n-type doping level is $n_0 = 4.9 \pm 1.0 \times 10^{12}$ cm$^{-2}$.

Other effects that may contribute to TR-ARPES with femtosecond EVU

In general, spectral shift of femtosecond EUV photoemission peaks can be possibly induced by pump surface photovoltage effect or space charging under intense pump excitation [3]. These two effects distort the movement of photoelectrons as they propagate to the detector, leading to shifts in the measured kinetic energy distribution. In surface photovoltage effect, the propagation of the photoemitted electrons in vacuum is affected by electric fields from pump induced dipoles in the sample [4–6]. For space charging, the photoelectrons are accelerated/decelerated by repulsive Coulomb interaction from photoelectrons generated via multiphoton ionization by the visible pump-pulse [7]. These two effects exist when pump and probe pulses are close in time, at both positive and negative delays, and will affect conduction band photoelectrons in quantitatively the same way as they do to valence band photoelectrons.

In our measurements, the conduction band energy distribution curve (EDC) remains constant within experimental uncertainty throughout the time delay window investigated, meanwhile the valence band EDC is up-shifted up at t>0. This indicates that the upshift of the valence band EDC after time zero is not from surface photovoltage effect or space charging. In addition, the space charge is strongly dependent on pump fluence. The applied pump fluence at 0.2 mJ/cm$^2$ is comparable or lower than that reported in the past TR-ARPES studies, in which space charging has not been observed [8–10].

The space charging effect emerges when the pump power is very high and induces a significant amount of photoelectrons by multiphoton ionization [3]. These photoelectrons will induce a

Coulomb repulsion to the probe photoelectrons, thereby shifting the photoemission signal to a different kinetic energy. The space charging effect will shift photoemission at both negative and positive time delays. In control experiments, we verify that the space charge effect indeed occurs at much high pump laser fluences (>> 1 mJ/cm$^2$) than those used in the TR-ARPES measurements on monolayer MoS2. As shown in Figure S7, the space charging effect induces a decay of the photoemission signal before and after time zero. It also shifts the CB and VB photoelectrons simultaneously.

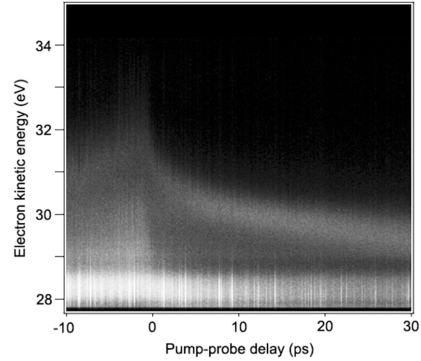

Figure S7. Space charging observed for monolayer MoS$_2$. The pump beam at 510 nm is focused very tightly on the sample to >> 1 mJ/cm$^2$, inducing a significant amount of photoelectrons. Note the curvature before and after time zero.

Photovoltage effect

For an n-doped monolayer MoS$_2$, an upward band bending in the lateral direction is expected at the junction with the Au electrode [11]. Across gap optical excitation by the pump pulse un-bends the band and this is equivalent to inducing a new dipole on the surface. As discussed before for the surface on a bulk 3D sample, the surface dipole layer will interact with the photoelectrons from the probe laser pulse [4]. At negative pump-probe delay (Δt < 0), the probe pulse arrives at a time delay -Δt before the pump pulse. In this case, the photoelectron will propagate during the time delay Δt for a certain distance z = vΔt, where v is the velocity of the photoelectrons emitted by the probe pulse. When the pump arrives at the surface and induces a dipole, the photo-electron will be susceptible to the electric field at distance $z = v\Delta t$ away from the sample. Therefore photoemission signal at negative time delays is correlated to the field intensities at different vertical distance z due to the visible pump-induced surface dipole.

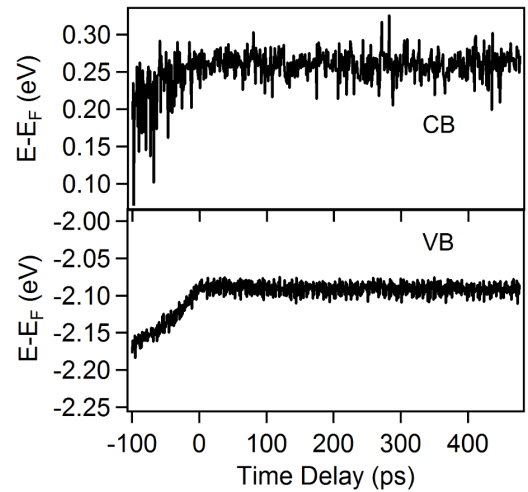

Figure S8. Long time scan of the CB and VB position as a function of visible pump EUV probe time delay. The shift before time zero is due to surface photovoltage effect, which simultaneously shifts the CB and VB positions. The time resolved trace before time zero maps the decay of electric field over space, while the time resolved trace after time zero corresponds to decaying of electric field over time.

At $\Delta t > 0$, the probe pulse arrives at a time delay $\Delta t$, after the pump pulse. Therefore the photoelectrons are affected by a pre-existing electric field induced by the pump pulse at $\Delta t$ near zero. In this case, the propagation of photoelectrons is affected by the electric field that has decayed during the time interval $\Delta t$. Thus, at $\Delta t > 0$, the photoemission signal is correlated with decaying surface dipole field at different delay times.

The surface photovoltage effect will affect both CB and VB signals simultaneously. Figure S8 shows a long time scan where the surface photovoltage effect is more prominent. The slope at negative time delay corresponds to the decay of pump induced electric field at different distances along vertical axis z. At positive time delays, the positions of the photoemission remains nearly constant, indicating that decay of the surface photovoltage at the experimental time window is very slow and negligible compared with the fast dynamics in the 0-80 ps window discussed in the main text. Since surface photovoltage effect shifts the position of both VB and CB simultaneously, it does not affect the measured band gaps.

Long-lived photo-excitation

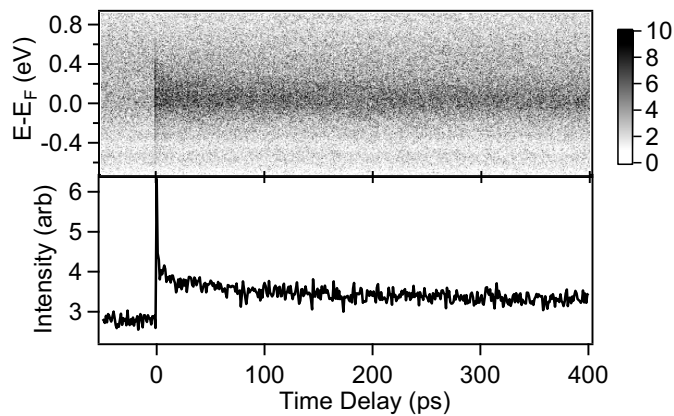

Figure S9. Conduction band photoelectron signal as a function of pump probe time delay from monolayer MoS$_2$, shown for pump-probe delay upto 400 ps. The photoexcited conduction band population is clearly observable at 400 ps. The initial excitation density is $n_{e/h} = 1.3 \times 10^{13}$ cm$^{-2}$ and sample is at 295 K.